\documentclass[aps,prl,twocolumn,amsmath,amssymb,showkeys,superscriptaddress]{revtex4-2}

\usepackage{graphicx,epsfig}
\usepackage{amssymb}
\usepackage{amsmath}
\usepackage{bm}
\usepackage{textcomp}
\usepackage{color}
\usepackage{float}

\newcommand\resxx{\mbox{Re($\sigma_{\rm xx} $)}}
\newcommand\sxx{\mbox{$\sigma_{\rm xx} $}}
\newcommand\fig[1]{Fig.~\ref{fig#1}}
\newcommand\Fig[1]{Figure \ref{fig#1}}

\newcommand\fp{\mbox{$f_{\rm p}$}}
\newcommand\red[1]{{#1}}
 
\begin{document}
\setcounter{page}{1}

\title[]{Origin of pinning disorder in magnetic-field-induced  Wigner solids}

\author{Matthew L. \surname{Freeman}}
\affiliation{National High Magnetic Field Laboratory, Florida State University, Tallahassee, Florida 32310, USA}

 \author{P. T. \surname{Madathil}}

\author{L. N. \surname{Pfeiffer}}
\author{K. W. \surname{Baldwin}}
 \author{Y. J. \surname{Chung}}
 \affiliation{Department of Electrical and Computer Engineering, Princeton University, Princeton, New Jersey 08544, USA}
\author{R. Winkler}
\affiliation{Northern Illinois University, DeKalb, Illinois 60115 USA}
\author{M. \surname{Shayegan}}
\affiliation{Department of Electrical and Computer Engineering, Princeton  University, Princeton, New Jersey 08544, USA}
 \author{L. W. \surname{Engel}}
 \affiliation{National High Magnetic Field Laboratory, Florida State University, Tallahassee, Florida 32310, USA}

\date{\today}
\begin{abstract}

 At  low Landau level filling factors ($\nu$),  Wigner  solid  phases of  two-dimensional electron systems  in GaAs are pinned by disorder, and exhibit  a pinning mode, whose frequency is a   measure of the disorder that pins the Wigner solid. Despite numerous studies spanning the last three decades, the origin of the disorder that causes the pinning and determines the pinning mode frequency remains unknown. Here we present a study of the pinning mode resonance in the low-$\nu$  Wigner solid phases of a series of  ultralow-disorder GaAs quantum wells which are similar except for their  varying well widths,  $d$.    The pinning mode frequencies, \fp,  decrease strongly as $d$  increases, with the widest well exhibiting \fp\ as low as $\simeq$35 MHz.    The amount of reduction of \fp\ with increasing $d$ can be explained remarkably well by   tails of the  wave function impinging into the alloy-disordered 
 Al$_x$Ga$_{1-x}$As barriers that contain the electrons. However, it is imperative that   the model for the  confinement  and wave function includes the Coulomb repulsion  in the growth direction between the electrons as they occupy the quantum well. 
  \end{abstract} 

 \maketitle

 When the   Coulomb  energy of  a system of electrons dominates  the kinetic energy, and disorder is sufficiently small, minimization of the Coulomb energy drives the formation of a  Wigner solid  (WS) \cite{wigner}, which is invariably pinned by interaction with the host system. 
Because of  pinning by disorder, unlike  the essentially uniform pinning of WS in some moir\'{e} systems \cite{shayeganflatband,padhi}, 
  a finite  correlation length of  crystalline order  is produced.  The pinning makes the WS an insulator, and also causes the   spectrum of a WS  to exhibit a pinning mode,  a small, collective oscillation of electrons of the solid about their pinned positions.  The frequency of the pinning mode is a measure of  the strength of the  pinning disorder.  In this Letter we utilize the pinning mode to identify  the disorder pinning a WS in state-of-the-art, ultralow-disorder GaAs quantum wells  \cite{chungultra,chungprb} in high magnetic fields.

There is a great deal of   evidence for the existence of WS in low-disorder, GaAs-hosted two-dimensional electron systems (2DESs)  at high magnetic fields, which effectively freeze out the kinetic energy \cite{lozo,lamgirvin,levesque,kunwc,msreview}. 
  These include pinning-mode studies   \cite{eva,williams91,paalanen,lessc,clidensity,yewc,yongab,yongmelt,zhwimbal,byoungdis,hanfqhwc,sciadv,yang},       $I$-$V$ curves \cite{goldman}
  higher-temperature transport \cite{panwcxtn}  and   photoluminescence \cite{goldys,buhmann,kukushkin}.    Most recently,  evidence for the WS has also come from  studies of the tunneling density of states \cite{ashoori},  from the geometric resonance of nearby mobile  carriers in a composite fermion metal by a WS in a nearby plane  \cite{mansourbi}, and from NMR relaxation \cite{tiemann}.    In  
  low-disorder GaAs 2DESs, a WS is expected in  high magnetic fields, starting near the fractional quantum Hall effect (FQHE) at Landau  level filling ($\nu=1/5$). Theoretical work \cite{rhim,archer}  has characterized the high magnetic field WS  to be  composed of   composite fermions  \cite{jainbook}, which can be thought of as electrons bound to an even number of flux quanta. 
The  present study of WS pinning disorder then has relevance  
to  the disorder that produces  vanishing diagonal conductivity in FQHE states, in which composite fermions are localized, either as single particles or as a pinned solid  \cite{hanfqhwc}.   A more detailed understanding of disorder effects on carriers  in high magnetic field may also be of value  to the  development of quantum computation schemes utilizing FQHE quasiparticles \cite{nayakreview}.  
  
Here we present a study of four GaAs quantum well (QW) samples,  grown by molecular beam epitaxy to be  similar to each other except for their well thickness, $d$. \red{ The samples  are grown according the state of the art described in Ref. \onlinecite{chungultra}. The QWs have   $d=30, 40, 50$ and $70$ nm,  and respectively mobilities 10,  17,  16 and    26 $\times 10^{6} $ cm$^2$/Vs and carrier densities of $n=4.5,4.5,4.2$ and $4.7\times 10^{10}$ cm$^{-2}$. While mobilities depend on $n$, the overall trend is of  mobility increasing with $d$.  According to the calculations of Ref. \onlinecite{chungprb}, with realistic estimates of the interfacial asperity characteristics,  interface  roughness disorder reduces the mobility for the narrower wafers.}   Flanking the QWs on each side are 280-nm-thick barriers of Al$_{0.12}$Ga$_{0.88}$As.
   In all four samples only the lowest  QW electric subband is populated. \red{The QW thicknesses and carrier densities of our samples are not compatible with a low $\nu$ bilayer (two-component) WS \cite{narasimhanho,haribi,hatkebi}.}
   
   We 
find a strong decrease in the pinning mode frequency, \fp,  on increasing $d$.  The measured   \fp\ decreases by nearly an order of magnitude as $d$ increases from 30 to 70 nm.  The 70-nm-wide  QW has an extremely low   $\fp\simeq35 $ MHz, indicating the extremely low disorder  of the sample. \fp\ vs  $d$ can be fit  very well using a model of alloy disorder arising from penetration of the tails of the wave functions into the   Al$_x$Ga$_{1-x}$As barriers that flank the well.  To explain  the data, the model must use a  Poisson-Schr\"{o}dinger approach that self-consistently incorporates  the  effect of the populated well on the charge distribution along the growth direction.

   
To measure the radio-frequency (rf) conductivity   we used coplanar waveguide (CPW) transmission lines patterned onto the top surfaces of the   samples as shown schematically in the inset of \fig{sbs}.  The CPW couples  capacitively to the 2DES,  about 900 nm below.  The driven center conductor of the CPW was 75 $\mu$m wide, and was separated from the grounded side planes by $ 50$-$\mu$m-wide slots. The transmission coefficients    were measured with a room-temperature network analyzer.   We calculated the 2DES conductivity, \sxx, from a distributed model of the coupled system of the CPW and 2DES. The model,  described in Ref. \onlinecite{zhwimbal}, was suggested originally in Ref. \onlinecite{foglerhuse}, and takes the 2DES to be in the small-wave-vector limit.  The model corrects for the tendency of the rf field in the 2DES to spread out from the region under the CPW slots near the resonance condition, particularly at low frequencies.   Pinning-mode frequencies,  from which the conclusions of this study are drawn,  are essentially the same whether assessed from the raw transmission coefficient spectra  or from the \resxx\ data shown below. 
 The  measurements were carried out in the low-power limit, in which the measured conductivity  is not sensitive to the excitation power, with the 2DES at $\simeq 35\,$mK, estimated from the dependence of the resonance on cryostat temperature.

  The experimentally deduced 2DES conductivity, \resxx, vs magnetic field, $B$, is shown in \fig{sbs} for the 30-nm QW, at a frequency of 250 MHz. Minima in \resxx\ due to FQHEs are clearly visible.   The   \resxx\ behavior between the 2/9 and 1/5 FQHE minima is an effect of  the varying proximity of the pinning resonance frequency to the measurement  frequency in that $\nu$-range.
  The apparent increase of \resxx\ as $B$  goes below 4 T comes from rf parallel conduction in the sample, which freezes out as $B$ increases. 

\Fig{allspx} shows color-scale   \resxx\  spectra taken at many fillings, $0.14<\nu<0.22$, for  the four QWs.
 For $\nu\lesssim 0.19$, all the QWs develop a striking, sharp resonance whose frequency, \fp,  depends strongly on the QW thickness.  Low \fp, observed particularly for the 70-nm QW, signifies  a long correlation length of crystalline order.  For the 70-nm QW the  number of electrons per (Larkin)  domain can be estimated \cite{chitra,foglerhuse}, as $2\pi\mu_c/eB\fp\simeq 1000$,  using the classical WS shear modulus \cite{bonsallmaradudin}, $\mu_c=0.245 e^2 n^{3/2}/4\pi \epsilon_0\epsilon$.
 Consistent with the interpretation of the resonance as a pinning mode, the resonance is absent in the range of the 1/5 FQHE, which   
 is clearly visible as a low-conductivity band that appears as dark blue.  
 As  1/5   is approached from either the  higher  or lower $\nu$  side, the resonance frequency, \fp, is reduced sharply as the resonance  fades away \cite{delecret}.   For $\nu$  above the 1/5 FQHE, in the reentrant  range \cite{reentrant}  of the WS,  up to $\nu=0.22$, a resonance is also present, although much weaker and broader.   The 40-nm QW data show a weak but sharp  secondary resonance at higher frequency than that of the main peak, although such a resonance is absent in the spectra of the other samples.  While likely related to the main peak, this secondary peak  does not appear to be a harmonic. 

\begin{figure}[t]
\includegraphics[width=\columnwidth]{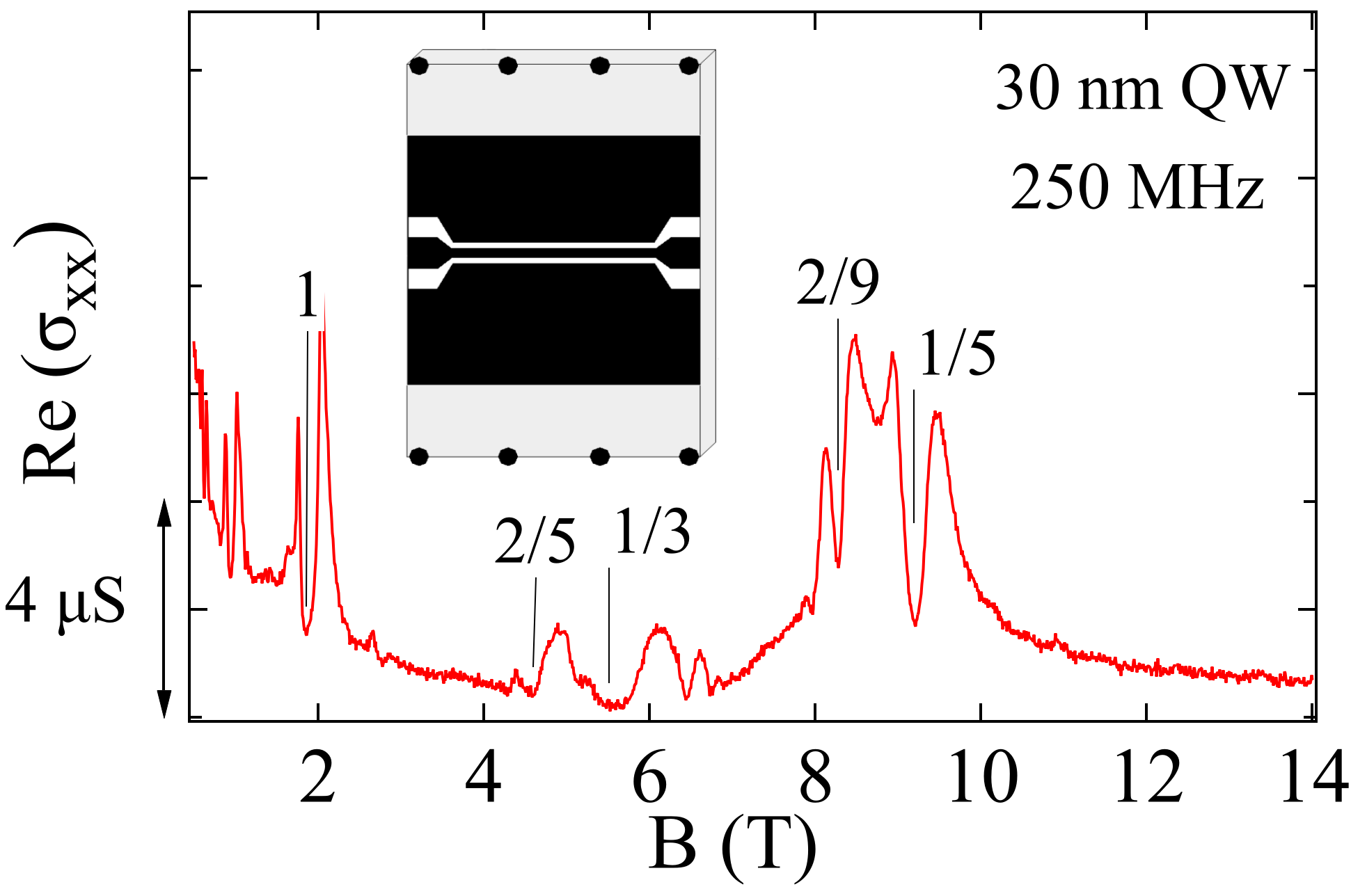}
\caption{     Real diagonal conductivity, \resxx, vs magnetic field, $B$ for the 30-nm QW. Inset: Schematic of measurement setup, with the metal film of coplanar waveguide shown as black. 
\label{figsbs}}
\end{figure}  

\begin{figure*}[t]
\includegraphics[width= \textwidth]{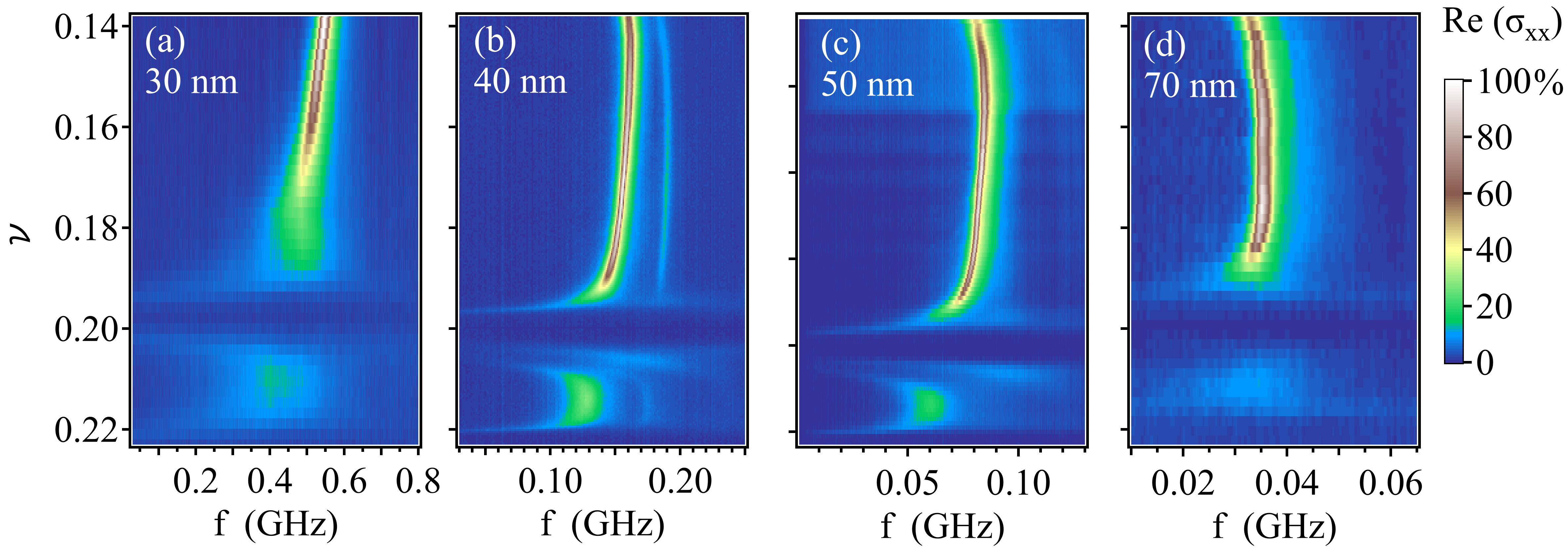}
\caption{Color-scale plots of   real diagonal conductivity, \resxx, on frequency-filling factor ($f$-$\nu$) axes, for the four QWs, marked with well widths, 
$d$. \red{ The graphs  are shown with larger $\nu$ at the bottom, so magnetic field, $B$, increases going upward.} The color scale is in percentages of full scales, which are 50, 210, 120 and 100 $\mu$S respectively for (a) through (d). For the 70-nm QW of panel (d) only, \resxx\ is multiplied by a factor of five for $\nu>0.20$, to enhance visibility in that range, which has small \resxx. 
\label{figallspx}}
\end{figure*}


The central result of this work is the pronounced decrease of \fp\ with increasing  QW thickness. 
 Though \fp\ is determined by   the shear modulus ($\mu$) of the WS  as well as by the pinning disorder \cite{chitra,fertig,foglerhuse}, we rule out the shear modulus effect as an explanation for the observed tenfold decrease   of \fp\ as $d$ is increased.   $\mu$  is expected to  decrease  as  $d$  increases, because in a thicker well the  electron-electron interaction is softer at short range.  However, in  the context 
 of weak pinning  \cite{chitra,fertig,foglerhuse}  a softer WS results in a {\em larger} \fp.  The inverse relationship of \fp\ and $\mu$ follows from the tendency of carriers, when the WS is softer,  to be more closely associated with disorder potential minima   and hence to experience larger 
restoring force upon displacement.

The densities, $n$, of our QWs are not identical, and $n$ can affect \fp, because $\mu$ increases with $n$. For well-developed resonances in the $\nu$ ranges shown here,  the expected relation is  $\fp\propto n^{-3/2}$.  This relationship is predicted in theory  \cite{fertig,chitra,foglerhuse} and  is also borne out by studies of \fp\ vs $n$ \cite{clidensity}.  To capture the 
trends solely due to changing $d$ in our \fp\ vs $d$ data we multiply the measured \fp\ by a near-unity   factor  $\alpha=(n/4.5\times 10^{10} {\rm\  cm}^{-2})^{3/2}$ to correct all the frequencies to their equivalents for  $n=4.5\times 10^{10} {\rm\  cm}^{-2}$. We emphasize  that the corrections are small: for  $d=30,40,50$ and $70$ nm QWs respectively, $\alpha=1.0,1.0,0.90$ and $1.07$. It is clear that the  variation of $n$ from sample to  sample cannot explain the large variation of \fp\ with $d$. 
 
  \Fig{fpkd} shows  $\alpha  \fp$  at $\nu=0.15$  
  vs  $d$. The graph is plotted in log-log format to account for the wide range of \fp\ as $d$ is varied.    The change of \fp\ with $\nu$ where the resonance is well-developed ($0.14\le\nu\le 0.19$) is shown as the error bars on the figure.     The other    curves in \fig{fpkd} show different model calculations for \fp\ vs $d$, and are discussed below.


 In the remainder of this Letter we consider different sources of disorder in our samples and show how the variation of $\alpha\fp$ with $d$
 points to the origin of the disorder that pins the WS.  
The types of disorder affecting the 2DES in  QWs are well-known:
(1) Charged impurities (usually C) which are dispersed through the growth, and   which  limit \cite{chungprb} the \red{possible} mobilities at zero magnetic field for dilute samples   with very thick spacer layers;
(2) Smooth disorder, stemming from the  granularity of the remote  donors grown into the sample, on a  long length scale given by  the  spacing of the donors; (3) Interface  roughness at the edges of the QW;  and finally (4) Alloy disorder \cite{byoungdis}, from  penetration of the tails of the    wave functions  into the   Al$_{\rm x}$Ga$_{\rm 1-x}$As  barriers, whose Al is randomly distributed.    The dispersed  impurities (1) would not be expected to produce the observed decreasing \fp\ with $d$.   Reference \onlinecite{chungprb}  shows that the effect of this disorder on mobility is independent of $d$.

  The remote ionized  donors (2) are about the same distance away from the WS for all the QWs, and also would not be expected to produce 
the strong thickness dependence.  This leaves interface roughness or alloy scattering in the barriers as candidates to explain the data.  

\begin{figure}[t]
\includegraphics[width= \columnwidth]{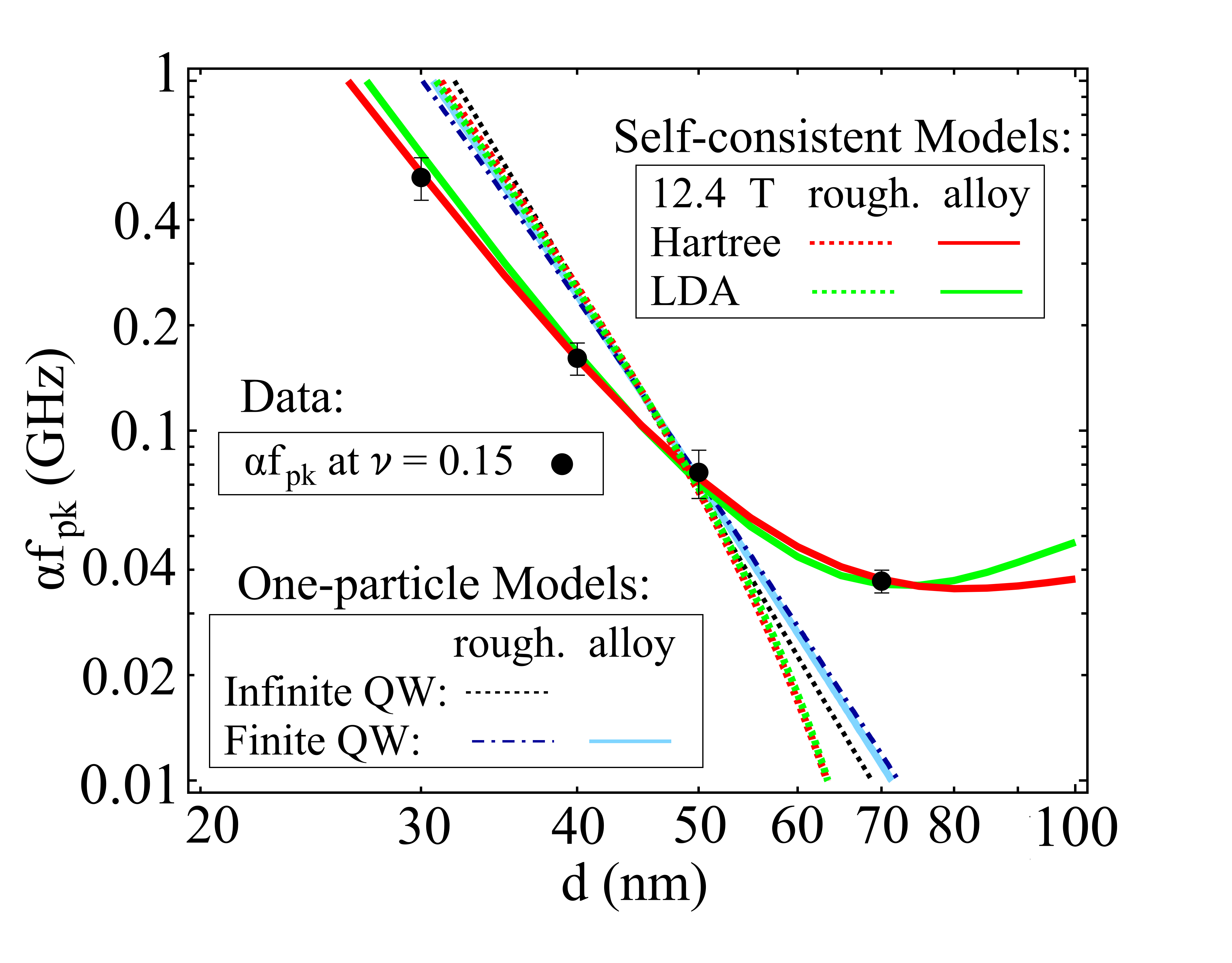}
\caption{  Density-corrected resonance frequency, $\alpha\fp$ (see text),   vs QW width, $d$.  Measured $\alpha\fp$ for the  four QWs at $\nu=0.15$ are shown as black dots. For  $d=30,40,50$ and $70$ nm respectively, $\alpha=1.0,1.0,0.90$ and $1.07$. 
Curves for the models discussed in the text are as shown in the legend. The one-particle   infinite QW (interface roughness) and finite QW (roughness and alloy) curves are multiplied by scale factors to intercept  $(d,\alpha\fp)=(45.3 $ nm$, 0.125 $ GHz$)$,   the centroid of the measured $\alpha\fp$ data points on the log-log plot. The curves marked ``alloy" are the calculated $P_B^2$, and  all ``roughness'' curves are the calculated    $(\partial E_s/\partial d)^2$; see text for details.  The $P_B^2$ curves generated for the Hartree and LDA self-consistent calculations are scaled to least squares fit the  $\alpha\fp$ data. The   fits are done on the logged   $\alpha\fp$ data.
\label{figfpkd}}
\end{figure}

\begin{figure}[t]
\includegraphics[width= \columnwidth]{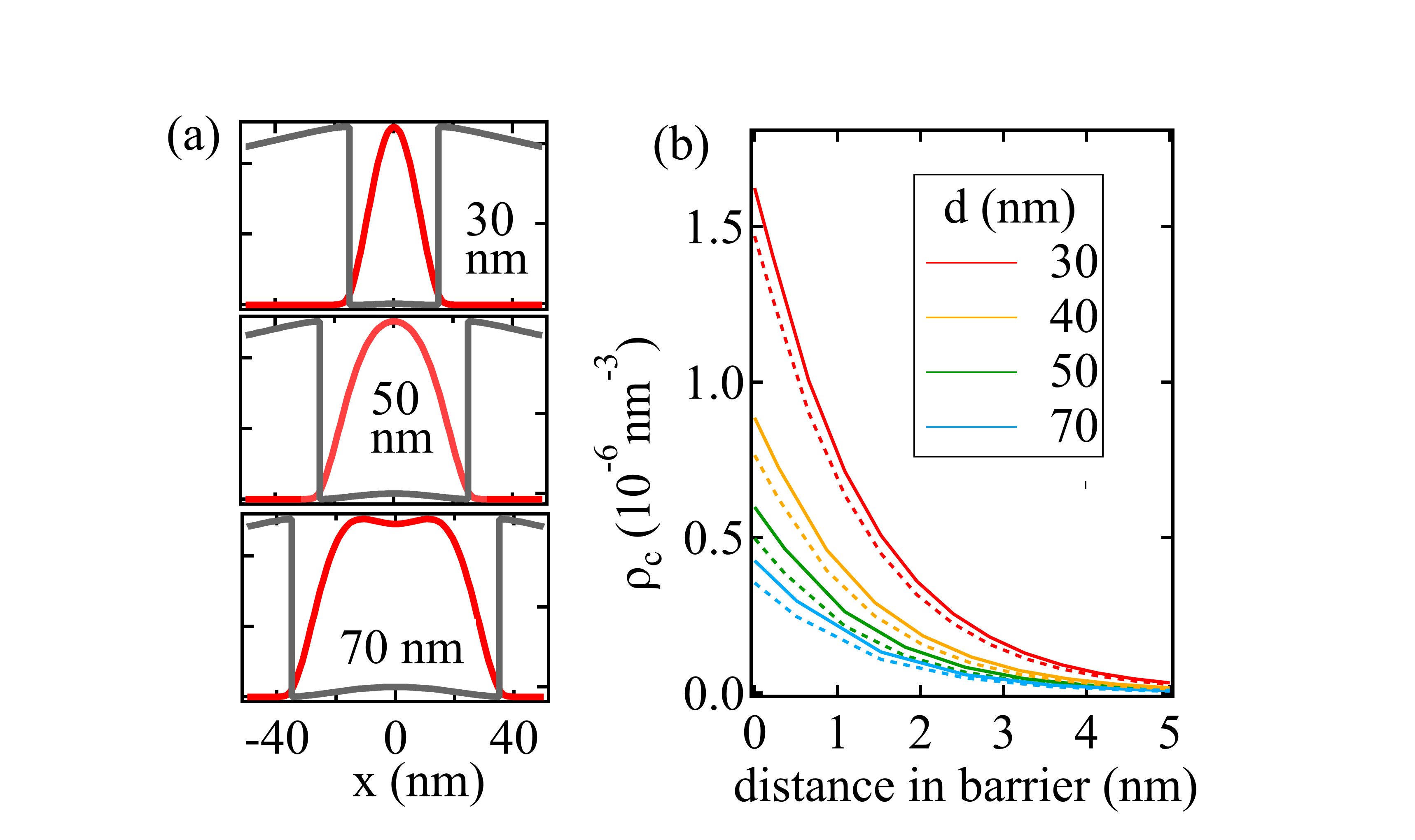}
\caption{  (a) Normalized charge density, $\rho_c$, (red), and self-consistent potential, $V$ (black) for QWs of  widths, $d= 30,50$ and $70 $ nm.   These are from  Poisson-Schr\"{o}dinger, self-consistent (Hartree) calculations for B=12.4 T.  (b) Hartree (solid) and LDA (dashed) calculated 
charge distribution tails within the barriers, for  the four QWs. The edge of the barrier is the origin of the horizontal axis.  
\label{figroland}}
\end{figure}

 To identify the pinning disorder,  we make use of  an {\em effective}  disorder potential, $V( {\bf r})$, to estimate the pinning energy.   The effective potential takes into account the probability that the electron wave function  in the WS will couple to a source of pinning disorder.   In weak-pinning theory  \cite{chitra,foglerhuse},  \fp\  goes as  a correlation function of this effective potential, 
 $\langle V(0)V({\bf r})\rangle =V_0^2 f(r)$, where $f(r)$ is normalized to unity and decays with length much shorter than the WS lattice constant,  so $\fp\propto V_0^2$ for a given type of disorder.     We  model $\fp(d)$ by considering the  dependence of $V_0^2$ on  $d$  to within a constant of proportionality, which we treat as an adjustable  parameter.

 Interface roughness disorder has been modeled based on the spatial fluctuations of the QW  subband energy   
  \cite{sakaki,kamburovrough,luhmanrough}. In this picture, asperities  in the interface locally alter the well thickness and hence the subband energy.    For an infinitely-deep QW, the subband energy, $E_s$, goes as $d^{-2}$, and a fluctuation in $d$ of $\xi \ll d$    produces a change  in $E_s$ of $  \sim -\xi d^{-3}$.  The interface roughness fluctuation size, $\xi$,  should be the same for all the samples, because they were grown in  the same system and under similar conditions.  So in considering the sample series,  and taking the fluctuation in subband energy to be $V_0$, we find  $ V_0^2\propto (\partial E_s/\partial d)^2\propto d^{-6}$.  We also obtained $E_s$ and $(\partial E_s/\partial d)^2$ from a simple model of a finite square well, whose depth is  the conduction band discontinuity, 0.095 eV  
\cite{adachi},  for the Al$_{0.12}$Ga$_{0.88}$As   barriers  closest to the QW, and effective mass of $0.067m_0$, both inside and outside the QW, where $m_0$ is the free-electron mass.      
The $(\partial E_s/\partial d)^2$ curves for the infinite and finite square wells are shown in \fig{fpkd} as a dotted gray line and a dashed-dotted   blue line. For comparison with the data the  curves are multiplied by   prefactors  to intersect the centroid of the data on the log-log plot.  The slopes of the two curves are similar, and are both  clearly   too large compared to the data.    
 
For effective alloy disorder   from the evanescent penetration  of the  WS charge density  into  the barriers,  $V_0$ is proportional to the  probability $P_B$,  that a carrier is within the barriers, so \fp\ is  modeled to be proportional to $P_B^2$.  In the simple, finite  square QW model of the previous paragraph,  $P_B$ is obtained by integrating the tails of the charge distribution  in the barriers.  The resulting $P_B^2$, again multiplied by a   factor to intersect the centroid of the   data,  is shown in \fig{fpkd}  as  a solid blue curve.  Clearly, this line also decreases with $d$ more strongly than the experimental  \fp\ data.

The simple ``textbook", one-particle,  square well models neglect  the effect of Coulomb repulsion  between the electrons along the  growth direction. This repulsion tends to push charge to the edges of a QW as it is widened \cite{suen91,suen92,hari96}.  We   performed 
 self-consistent calculations  of the charge distribution and potential, using the Poisson and Schr\"{o}dinger equations, at magnetic field $B=12.4$ T, and 
 $n=4.5\times 10^{10}$ cm$^{-2}$ to  match $\nu=0.15$ as in \fig{fpkd}. 
  \Fig{roland}(a)
 shows   potential and charge distributions calculated for  the   30, 50 and 70 nm QWs in the self-consistent Hartree approximation \cite{rolandbook}. As $d$ increases,  the charge distributions flatten and eventually become bimodal.

 We carried out the  self-consistent calculations for high magnetic field, with the charge density coming from the occupied Landau level only. Because of the 
 Landau level degeneracy, the second electric subband, occupied for $d>70$ nm at $B=0$, is not occupied in high field. We evaluated the wave functions of the Landau levels via an $8 \times 8$ multiband Hamiltonian \cite{rolandbook}. Besides the self-consistent Hartree approximation, we also  used 
 spin density-functional theory in the local-density approximation  (LDA) based
on the spin-dependent exchange-correlation potential parameterized
in Ref. \cite{perdew}.

Estimates of the effect of interface roughness  in the context of the self-consistent calculations come  from taking $\fp\propto(\partial E_s/\partial d)^2 $, with $E_s(d)$ (measured from the well center) coming from the self-consistent calculations. The Hartree and LDA calculations give  the red and green  dotted curves in \fig{fpkd}, respectively.  The curves  are again multiplied by scale factors to intercept the centroid of the measured data.   The curves are nearly   identical and    decrease much more strongly than the experimental data.

 In  \fig{fpkd},   the red and green  solid curves   show   $P_B^2$ vs $d$ obtained from the Hartree and LDA calculations respectively, 
 multiplied by  scale factor parameters to least-squares fit the density-adjusted experimental data, $\alpha \fp$.   The fits are remarkably good.  $P_B^2$  is proportional to the contribution of alloy disorder to \fp,  so the good fits 
  suggest that the charge density  in the barriers is indeed the origin of the pinning disorder, and that  the self-consistent modeling of the populated well is required to reasonably estimate the $P_B$ associated with the charge density in the barriers. In comparison to the one-particle square-well models, the  self-consistent models lessen  the effect of the widening well, because the pushing outward   of the charge distributions from  Coulomb  repulsion preserves more charge density in the barriers as the well gets wider.  \red{In the theoretical curves, this effect continues at larger $d$, so that the curves increase 
  for $d> 70$ nm. }

The self-consistent  models are only  approximations because they take the charge as uniform sheets in the 2D plane, which is not the case in a WS. For our samples with  $n\simeq 4.5\times 10^{10}$ cm$^{-2}$, the effective in-plane size of an electron is the magnetic length $l_B\simeq7$ nm at $\nu=0.15$, while the  WS lattice constant $a=2^{1/2}\cdot 3^{-1/4}n^{-1/2}\simeq50$ nm.   In between WS lattice sites, because of  the smaller  charge density,  the penetration of charge into  the barriers  will be somewhat less than the model estimates.  Likewise more penetration than predicted will occur  at lattice sites.  These  effects would not vary much with $d$, explaining the strikingly good fits we see.


In summary, we have studied the pinning mode of the  magnetic-field-induced  WS in a series of ultralow-disorder GaAs  QW samples, of varying thickness. An extremely low  \fp\ down to 35 MHz is found in the 70 nm QW.   We identify the disorder responsible for the WS pinning as  Al alloy disorder in the QW barriers.    For  this disorder, the decreasing curve  of \fp\ vs well thickness is quantitatively explained  by  the reduction of charge density in the barriers for thicker wells, if this reduction is  self-consistently evaluated to account for the QW population. 
 


%
%
%
%
%
%
%

 \begin{acknowledgments}

 A portion of this work was performed at the National High Magnetic Field Laboratory (NHMFL), which is supported by National Science Foundation (NSF) Cooperative Agreement No. DMR-1644779 and the state of Florida. The rf measurements   at NHMFL were supported by Department of Energy (Grant No. DE- FG02-05-ER46212).  Work at Princeton University was supported by the NSF Grant No. DMR 2104771 for measurements, by the U.S. Department of Energy Basic Energy Office of Science, Basic Energy Sciences (Grant No. DEFG02-00-ER45841) for sample characterization, and the NSF Grant No. ECCS 1906253 and the Gordon and Betty Moore Foundation's EPiQS Initiative (Grant No. GBMF9615 to L.N.P.) for sample synthesis. We also thank David Huse,  J. K. Jain and Adbhut Gupta for illuminating discussions.
\end{acknowledgments}

\bibliography{qws.bib}
 

\end{document}